\documentstyle[proc]{rspublic} \input epsf.tex
\def\bi{\begin{itemize}} \def\ei{\end{itemize}}
\def\bq{\begin{quotation}} \def\bra{\langle} \def\ket{\rangle}

\def\eq{\end{quotation}}

\def\thedemobiblio#1{\smallskip\par
 \list{}{\labelwidth 0pt \leftmargin 1em \itemindent -1em \itemsep 1pt}
 \small \parindent 0pt
 \parskip 1.5pt plus .1pt\relax
 \def\newblock{\hskip .11em plus .33em minus .07em}
 \sloppy\clubpenalty4000\widowpenalty4000
 \sfcode`\.=1000\relax}

\begin{document}

\title[Dynamical mean-field approach]{
Dynamical mean-field studies of  metal-insulator 
transitions }

\author[Dobrosavljevi\'{c} \& Kotliar]{V. Dobrosavljevi\'{c}$^1$
and Gabriel Kotliar$^2$}

\affiliation{$^1$Department of Physics and 
National High Magnetic Field Laboratory,\\Florida State University, 
Tallahassee, Florida 32306, USA\\
$^2$Serin Physics Laboratory,
Rutgers University, \\PO Box 849, Piscataway, NJ 08855, USA}
\maketitle
\begin{abstract}
We discuss the successes of the dynamical mean field (DMF) approach to metal
insulator transitions in both the clean and the disordered limit. In
the latter case, standard DMF equations are generalized in order to
incorporate both the physics of strong correlation and Anderson
localization effects. The results give new insights into the 
puzzling features of doped semiconductors.
\end{abstract}


\section{Introduction}

How a substance evolves from  a metallic to a non metallic
state is one of the most fundamental
and richest  problems in condensed matter physics.

In general, there are several mechanisms at play.  
Electron-electron interactions can drive a metal to insulator
transition in a pure substance (Mott, 1990). This transition is named after Sir
Nevill Mott who laid down the foundations for the physical
understanding of this phenomenon. Another route was discovered by
Anderson (1958) who realized that sufficiently strong
disorder can drive a metal-insulator transition even in systems of non
interacting electrons.  The theoretical description of the situation
when both effects are present is a central unsolved problem.  For
recent reviews see Lee \& Ramakrishnan (1983) and Belitz \& Kirkpatrick (1995).

Early treatments (Abrahams et al. 1979, Wegner 1976,
Sch\"{a}ffer \& Wegner 1980,  Finkelshtein 1983, 1984) used
direct analogies from the theory of magnetism and the scaling approach
to critical phenomena.  The technical apparatus of the approach was a
field theoretical non linear sigma model and an expansion near two
dimensions. The physical content of such theories was expressed
concisely in terms of an extension (Castellani et al. 1987) of the
Fermi liquid approach to disordered systems.  These ideas were very
successful in the description of the transport and thermodynamic
properties of weakly disordered metals.  The approach, however,
encounter several difficulties in accounting for experimental
observations in many systems. The origin of these difficulties can be
traced to (1) strong interaction effects, and (2) strong statistical
fluctuations.  One important manifestation of this effect was
the formation of local magnetic moments, objects which can not be
understood from a weak correlation perspective.  This aspect of the
problem and its theoretical treatment were addressed in Paalanen et
al. (1988) and Milovanovi\'{c} et al. (1989).

In the last few years, a new dynamical mean field approach to the
strong correlation problem has been developed (Metzner \& Vollhardt
1989, Georges \& Kotliar 1992).  It has explained many puzzling
features of clean three dimensional transition metal oxides (for a
review see Georges et. al. 1996).  Very recently, it was extended
(Dobrosavljevi\'{c} \& Kotliar 1997) to incorporate the interplay of
Anderson localization and correlations effects.

The mean field method is largely complementary to the scaling approach.
While the latter concentrates on the long wavelength modes which
presumably govern the immediate vicinity of the transition, the former
focuses on the local charge and spin dynamics of the electrons.  The
latter is based on an expansion around the lower critical dimension,
the former being formally valid in large dimensions.  It is our view
that a good mean field understanding of the metal to insulator
transition problem is a necessary first step towards a comprehensive
theory.

In this talk we will not focus on the technical aspects of the method
which are thoroughly reviewed elsewhere.  Instead we will attempt to
give a simple description of the physical content of the approach, and
of the results obtained by this method.  Section \ref{clean} introduces
the
essential idea of the mean field method for a general
periodic solid.  It is our hope that its
generality and local character will be of interested to the chemistry
community which has stressed the importance of local bonding and
correlations through the years.  Section \ref{statistical} discussed
the extension to the disordered case.  We stress the importance of
statistical fluctuations and how they are captured by the
mean field  method.  In
section \ref{mott} we discuss the physics of the density driven Mott
transition and explain  why the mean field approach works so well for
three dimensional compounds, stressing the role of orbital degeneracy.
We then move  to  the physics of disordered
interacting systems which was the focus of many years of Mott's
research.  Section \ref{puzzle} discusses our views on the puzzle
posed by the differences between  uncompensated semiconductors and
disordered alloys.  In section 5, \ref{new} and 7  we elaborate on the physical
content of the mean field approach and argue it captures some essential
elements needed to resolve this puzzle.
We conclude in section \ref{conclusions}
with a comparison to other approaches and outline some directions for
further work.

\section{Dynamical mean-field theory (DMF) theory - the clean limit}
\label{clean}

To motivate the dynamical mean field (DMF) approach, it is useful to
draw some analogies with a well established approach to the electronic
structure problem, the density functional theory (DFT). The basic
physical quantity in DFT is the density $\rho (r) $, and the free
energy of a system is written as a functional of the density
$F_{DFT}[\rho (r)]$.  Its minimum gives the physical density of the
system in question.  In practice, the form of the functional $F_{DFT}$
is unknown and various approximations such as the LDA (local density
approximation) are used instead of the exact density functional.

The dynamical mean field approach is very similar in spirit, except
that it adopts the local spectral function $A(\omega, r) \equiv -{1
\over \pi} Im G(r,r;  \omega + i \delta )$ as the basic quantity.
One then writes the free energy as a functional of this quantity, and
its minimization leads to the dynamical mean field equations which
give the one particle spectral function for the system of interest.
One can motivate the extension from the density to the spectral
function (which can be thought of loosely as a kind of energy-resolved
density ) in the context of strongly correlated electron
systems, by thinking about their photoemission (and inverse
photoemission) spectra which demonstrate the existence of
bands which have the character of atomic configurations (Hubbard
bands), in addition to the ordinary  quasiparticle bands  which
are analogous to those of the non interacting system.
One can incorporate quasiparticle bands and Hubbard bands
naturally in a single theoretical framework by  resolving  in
energy, the local density, i.e. the local density is the integral
of the local spectral function over frequency. 
By resolving in energy the localized and itinerant component
of the electron,   DMF approach treats
coherent (quasiparticle like) and incoherent (Hubbard bands like)
excitations on the same footing. It is a unified  framework for the
description of
localized and itinerant electrons.
 
To illustrate the generality of
the method let us start from a Hamiltonian containing several orbitals
per unit cell. We use a compact notation where
the  index
 $\alpha=(m,\sigma)$
 combines the    orbital $m$ and the  spin  $\sigma$.
\begin{equation} \label{e1}
H_{lattice}= -
    \sum_{\bra ij \ket } c^+_{i\alpha}\,t_{i \alpha,j\beta}\,
c_{j \beta}+ \sum_i \left(E_{\alpha \beta}-\mu \delta_{\alpha
\beta} \right) c^+_{i\alpha} c_{i\beta}
 + \sum_i \Gamma_{\alpha \beta \gamma \delta}\,
c^+_{i\alpha} c_{i\beta} c^+_{i\gamma} c_{i\delta}.
\end{equation}

We now focus on a single unit cell, and integrate out
all degrees of freedom except for those which reside
in the selected unit cell. These
are described by operators $c_{\alpha}$ and no
longer carry a site index.
The dynamics of the resulting problem is described
by
an impurity model which describes an impurity ($c_{\alpha}$\/)
coupled to a bath of fermions ($a_{b\mu}$\/) 
\begin{eqnarray} \label{e2}
H_{imp}& =& \sum_{\alpha \beta} \left(E_{\alpha\beta}-
 \mu\,\delta_{\alpha \beta}\right)\, c^+_\alpha c_\beta +
\sum_i \Gamma_{\alpha \beta \gamma \delta}\, c^+_{\alpha}
c_{\beta} c^+_{\gamma} c_{\delta}+
\sum_{b\mu} \epsilon_{b\mu}\, a^+_{b\mu} a_{b\mu}\nonumber \nopagebreak\\
& &+ \sum_{k} \left( V_{b\mu,\alpha}\, a^+_{b\mu} c_{\alpha} + h.c.
\right)
\end{eqnarray}

{From }the impurity model we can obtain
all the local  correlation functions, since
by construction the local lattice Green's functions
are identical to the impurity Green's function
 $\hat{G}=(G_{\alpha,\beta})$ .
We use a matrix notation so that the local Green's function
is given by
\begin{equation}
G_{\alpha,\beta}(\tau-\tau^\prime)=-\bra
T_\tau\, c_{\alpha}(\tau) c^+_{\beta}(\tau^\prime) \ket
\end{equation}

This should be viewed as a functional of the parameters
$ \epsilon_{b\mu}$ and $ V_{b\mu,\alpha}$.
To determine these parameters we construct the
``Weiss field" which describes the effect of the rest of the
electrons on the selected cell,
$\hat{G}_0^{-1}(i\omega_n)=(i\omega_n+\mu)\,\hat{I}-\hat{E}-\left(
\sum_{b\mu}\frac{V_{b\mu,\alpha} \,V_{b\mu,\beta}}{i\omega_n
-\epsilon_{b\mu}}\right)$, and the self energy of the impurity,
 $\hat{\Sigma}(i\omega_n)=
\hat{G}^{-1}_0(i\omega_n) - \hat{G}^{-1}(i\omega_n)$,
viewed as a functional of  $\epsilon_{b\mu}$ and
$V_{b\mu}$.
These parameters are determined by requiring that the bath and the
local degrees of freedom describe the electrons in the original
lattice problem. Namely we can construct the local Green's function
from the lattice Green's function obtained by adding a k independent
self energy to the non interacting lattice Green's function (obtained
from \ref{e1} by setting the interaction terms to zero) or from the
impurity model.

\begin{equation} \label{e3}
 \hat{G}(i\omega_n)=\sum_k \left((i\omega_n+\mu)\hat{I}- \hat{t}(k)
- \hat{\Sigma}(i\omega_n)-\hat{E}\right)^{-1}
\end{equation}

The reader will notice many common features between DMFT and the Bragg
Williams theory of magnetism. However in fermionic problems a new
feature emerges--the cavity field acquires nontrivial {\em time
dependence}, allowing a nonperturbative treatment of {\em local
dynamics}, which proves to be of crucial importance for strongly
correlated electrons. In particular, the approach incorporates {\em
incoherent} (inelastic) processes even on this mean-field level, as
opposed to most other treatments. As a result, the formulation can be
used even in the study of {\em non-Fermi liquid} metallic phases, for
example in extended Hubbard models (Si and Kotliar 1993).

\section{Dynamical Mean Field Theory of the Density Driven
Mott Transition in Pure Systems}
\label{mott}

The Mott transition in transition metal
oxides has received renewed theoretical and experimental attention.
On the experimental side, new compounds have been synthesized
(Tokura et al. 1993, Okimoto et al. 1995)
and known compounds such as $V_2O_3$ and $Ni Se_{1-x} S_x$ have been studied.
On the theoretical side, new  insights have been obtained
from studying the one band Hubbard model
in the limit of large lattice coordination.
In this limit the pressure and temperature
driven Mott transition as well as the density driven transition
can be thoroughly analyzed. 

Several quantitative comparisons between the physics of three
dimensional transition metal oxides and the one band Hubbard model
have already been performed (Kotliar \& Kajueter 1996, Rozenberg et
al. 1995).  For example, the doping dependence of the electronic
specific heat, the resistivity and the Hall coefficient in $La_{x}
Sr_{1-x} Ti O_3$ can be explained by the one band Hubbard model 
without adjustable parameters after the values of $U$ and $D$ have
been extracted from photoemission data (Kajueter et. al. 1996).
Similarly, the single band
Hubbard model can describe the temperature dependence of both, the
optical and the DC-conductivity in $V_2 O_3$ (Rozenberg et al. 1995).

The Mott transition with integer occupation has been
the subject of extensive reviews (Georges et al. 1996). Here we focus
on some qualitative aspects of the physics of the density driven Mott
transition, the emphasis will be on the question: why does the mean field
theory work so well for three dimensional transition metal oxides? 

We start with a brief discussion of the qualitative content
of the mean field theory of the doped Mott insulator.
At low temperatures the mean field theory describes
a Fermi liquid  with a an energy scale (renormalized
Fermi energy  ${\epsilon_F} ^*$) which is proportional to the
 doping (i.e. deviation from half filling)  $\delta$.
In infinite dimensions, the renormalized Fermi energy is
proportional to the quasiparticle residue Z which also
vanishes linearly in doping$, i. e. {\epsilon_F} ^* = ZD$ .
In this regime, the linear term of the specific heat
is inversely proportional to $\delta$ and  the Hall coefficient
is  unrenormalized from its band structure value.
The Fermi liquid description is valid up to a temperature
scale $T_0$ that surprisingly  scales as $\delta ^{3 \over 2}$,
and is therefore quite small very close to the  Mott transition
(Kajueter et. al. 1996). 
 
The qualitative success of the mean field approach when
applied to three dimensional transition metal oxides
is due to  the orbital degeneracy
of three dimensional transition metal oxides.
Orbital degeneracy causes a
close  competition between ferromagnetic and antiferromagnetic
tendencies. The net spin-spin interaction is a sum of
antiferromagnetic and ferromagnetic terms which tend to largely cancel.
An illuminating example that can be studied exactly is a
two site system was described by Kajueter and Kotliar (1997) mimicking
a Ti ion in $LaTi O_3$. 
The basic scale of the problem is a bandwidth which is of order of 1
eV, but  the splitting between different spin  configurations turns out to 
be a much  smaller ( between $10^-{3}$ and $10^{-4}$ eV).

If we insist on using a one band model which ignores the orbital
degeneracy, to reproduce the gross features of transition metal
oxides, one must introduce in the model terms that suppress the tendency
towards  magnetic order.  In infinite dimension,  this can
be accomplished  by choosing 
lattices with longer range hopping matrix elements which induce
magnetic frustration or by adding and additional ferromagnetic
interactions  to the
Hamiltonian to compensate for the antiferromagnetic exchange
which is inherent to the Hubbard model on the  hyper-cubic lattice.
A Hamiltonian of the form:
\begin{equation}
H = - \sum_{ij,\sigma} t_{ij}
c^{+}_{i\sigma} c_{j\sigma} +
U\sum_{i} n_{i\uparrow} n_{i\downarrow}+ 
{{J}\over{d}} \sum_{<ij>} \vec{S}_i\cdot\vec{S}_j
\label{tj}
\end{equation}
with an additional ferromagnetic interaction J 
increasing  as a function of x, can serve as a crude
caricature of  
the $La_{1-x}{ Y}_{x} Ti O_3$ system.
This compound is ferromagnetic for  x  near 1 but
becomes antiferromangetic at small values of x.  (Okimoto  et. al. 1995).

While this approach is certainly too  crude to describe
details of the magnetic ordering, it  expresses the
fact that the localization tendencies in this system  are
not caused by, and are
largely independent of,  the magnetic ordering (which can
be either ferromagnetic  or antiferromagnetic).
In the limit of large dimensions the
J term does not affect the one particle properties, 
so the calculations of the effective mass  performed
in the absence of this term apply to this model as well.
The Hamiltonian (\ref{tj} )   was  used 
by A. Georges \& L. Laloux (1996) to model  the Wilson
ratio of liquid $ He_3$, and by Obermeier et. al. (1996)
to model the magnetic response of cuprate superconductors.

To summarize, the mean field theory works very well
in systems where the charge and spin fluctuations are
reasonably local. This is the situation
in  three dimensional transition metal oxides
(as a result of orbital degeneracy) and in  the disordered
systems  described in the next section.

While the qualitative success of dynamical mean field theory is easy
to understand, the surprising quantitative agreement of this approach
as applied to orbital degenerate systems was clarified only recently(
Kotliar \& Kajueter 1996, Kajueter \& Kotliar 1997).  It was shown
that:  the quantitative value of physical quantities such as the
effective mass, and the optical gap, near the Mott insulating state
with {\it one electron per site} which had previously been compared
with experiments depend weakly on the band degeneracy (the differences
are of less than 10 percent). On the other hand  
the high energy behavior of the
spectral functions have a sizable dependence on the number of orbitals
per site.

\section{The metal to insulator transition  in  disordered systems: some outstanding puzzles	}
\label{puzzle}

The presence of disorder adds a new dimension to the metal to
insulator transition problem.  The continuous nature of the metal to
insulator transition was first demonstrated in this class of systems (
for a general review of metal insulator transitions see Mott (1990)).
As a result of a series of intensive experimental studies a basic
picture of the behavior of physical quantities near the transition has
emerged.  In this section we describe briefly some aspects of the
experimental picture which we regard as well established,  but which
still remain  a challenge for the microscopic theory.  For more
comprehensive reviews of the experimental situation in this field see
Paalanen \& Bhatt (1991)  Sarachik (1995)  and v. Lohneyesen (1997).

It is well established the $T=0$ conductivity vanishes as the
transition (i.e. critical dopant concentration in doped
semiconductors) is approached from the metallic side.  We leave the
immediate vicinity of the metal to insulator transition out of this
analysis because the extrapolation to zero temperature is problematic
(for a good discussion of this point see v. L\`{o}hneysen (1997)).  In
the region where this extrapolation is unambiguous, the conductivity
vanishes in a power law fashion.  $\sigma
\approx (n-n_c)^\mu$.  In uncompensated materials (one electron or
hole per dopant ion), the conductivity exponent takes an anomalously
small value $\mu\approx 1/2$ (Paalanen et al. 1980). Such a small
value of the conductivity exponent proved to be notoriously difficult
to explain by any known theory.  On the other hand, well compensated
doped semiconductors (only partial filling of the impurity band) 
 behave very differently; their conductivity vanishes
with an exponent $\mu\approx 1$.  Amorphous alloys behave the same
way.  In the presence of a strong magnetic field the conductivity
vanishes with an exponent $\mu\approx 1$, in both compensated an
uncompensated semiconductors (Dai et al. 1991, 1992).  The presence
of spin-orbit (SO) coupling (Dai et al. 1991, 1992) does not seem to
be a determining factor for the behavior of the conductivity.  Both
Si:B (where SO coupling is strong) and Si:P (where SO coupling is
weak) behave  qualitatively in the same way.  Uncompensated
doped semiconductors also display an anomalous temperature dependence.
In the metallic phase 
for  concentrations not too close to the transition
the conductivity {\em increases} as the temperature
is lowered. 
This
behavior can be reversed by applying strong magnetic fields, resulting
in a decreasing conductivity at low T.

In our view these anomalies occur over a wide range of concentrations
and are {\em not} restricted to a tiny critical region in the vicinity
of the transition.  For example, the anomalously small value of the
conductivity exponent $\mu\approx 1/2$ describes the data all the way
to $n\approx 4n_c$ ! It is then natural to conclude that the observed
behavior should not be identified with an asymptotic critical behavior
associated with a narrow critical region. Rather, it should be
described by an appropriate {\em equation of state}, that is expected
to follow from a relevant {\em mean-field} description of the problem.
We also emphasize that the anomalies are associated with {\em
uncompensated} systems, where we expect the effects of electronic
correlation to be the strongest. It is thus natural to approach the
problem from a DMF perspective. In this picture, many of the above
features, but in particular the compensation (filling) dependence
follow naturally.

In contrast to the  transport properties, the thermodynamic quantities
$\chi$ and $\gamma$ vary smoothly (Paalanen et al. 1988) as a function of
concentration across the transition, and seem to diverge at low
temperatures both on the insulating and on the metallic side of the
transition.  The NMR experiments (Paalanen et al. 1985, 
Alloul \& Dellouve 1987) portray a strongly
inhomogeneous picture.  There is a wide distributions of Knight shifts
on the Phosphorus sites.  As the transition is approached a large
number of sites acquire knight shifts that are larger than the
measurable range indicating the formation of local moments.  The
knight shift on Si is a smoother function of
concentration( Stoddart et al. 1992) suggesting that the metal insulator
transition takes place in the phosphorus impurity band.

Underlying this experimental picture is a broad distribution of energy
scales which makes the metal-insulator transition in disordered systems very
different than in the clean case. In addition, this behavior reflects 
unusually developed {\em spatial fluctuations} of the electronic system.
While such behavior does not normally occur in ordinary metals, it
is a natural consequence of the incipient {\em localization} of the 
electronic states -- which by definition cannot happen in a uniform fashion. 

\section{Statistical DMF theory}
\label{statistical}

We have argued that in presence of disorder, the situation is
qualitatively more complex than in the clean limit. 
Clearly, it is necessary to  extend
the DMF ideas in order to capture the crucial effects of disorder --
the spatial fluctuations of the order parameter.  In the following we
present a simple pedagogical derivation of the statistical DMF theory,
and discuss some of its main features.

We follow an approach very similar to the 
Thouless-Anderson-Palmer (1977) formulation of the mean field theory
of spin glasses.  Specifically, we treat the correlation aspects of
the problem in a dynamical mean-field theory fashion, but allow {\em
spatial variations} of the order parameter in order to allow for
Anderson localization effects.  The theory is then exact in the
noninteracting limit, and reduces to the standard dynamical mean field
theory in absence of disorder.  The approach can be applied to any
lattice model of interacting fermions.  For simplicity, we consider a
simple single-band Hubbard model with random site energies given by
the Hamiltonian


\begin{equation}
H=\sum_{ij}\sum_{\sigma } ( -t_{ij} + \varepsilon_i \delta_{ij})
c^{\dagger}_{i,\sigma}c_{j,\sigma} +
U\sum_{i}c^{\dagger}_{i,\uparrow}c_{i,\uparrow}
c^{\dagger}_{i,\downarrow}c_{i,\downarrow}.
\end{equation}

Following the general spirit of the DMF theory, we focus on a
particular site $i$ of the lattice, and integrate all the other
sites. This procedure is formally exact, but the resulting local
effective action takes an arbitrarily complicated form (Dobrosavljevi\'{c}
\& Kotliar 1994), containing n-point vertices of all orders. 
However, within the DMF approach, one retains only the 
contributions {\em quadratic} in Fermion fields, and the local effective 
action assumes the form 

\begin{eqnarray}
&&S_{eff} (i) =  \sum_{\sigma}\int_o^{\beta }d\tau\int_o^{\beta }d\tau '
c^{\dagger}_{i,\sigma} (\tau )( \delta (\tau -\tau ')\left(
\partial_{\tau} +\varepsilon_i -\mu \right) \nonumber \\
&& +  \Delta_{i,\sigma} (\tau,\tau ') )c_{i,\sigma} (\tau ')  
  + U\int_o^{\beta}d\tau n_{i,\uparrow}(\tau )n_{i,\downarrow}(\tau ).
\end{eqnarray}
Here, we have used functional integration over Grassmann  fields 
$c_{i,\sigma} (\tau )$ that represent electrons of spin
$\sigma$ on site $i$, and $n_{i,\sigma}(\tau )= 
c^{\dagger}_{i,\sigma} (\tau )c_{i,\sigma} (\tau )$.
The ``{\em hybridization function}''
 $\Delta_{i} (\tau,\tau ')$ is given by
\begin{equation}
\Delta_{i} (\omega_n ) =\sum_{j,k=1}^z t_{ij}^2 G_{jk}^{(i)} (\omega_n ).
\end{equation}
The sums over $j$ and $k$ runs over the $z$ neighbors of the site $i$, and

\begin{equation}
G_{jk}^{(i)} (\omega_n )=<c^{\dagger}_{j}(\omega_n )c_{k} (\omega_n )>^{(i)} 
\end{equation}

are the lattice Green's functions evaluated with the site $i$ removed. 
In general, these objects can be expressed through ordinary lattice Green's 
function as

\begin{equation}
G_{jk}^{(i)} = G_{jk} -\frac{G_{ji}G_{jk}}{G_{ii}}.
\end{equation}

We emphasize that the above construction is carried out for a {\em 
fixed realization} of disorder defined by a given set
of random site energies $\{ \varepsilon_i \}$. The DMF truncation, 
which keeps only the quadratic contributions to the effective action is 
{\em exact} for:\vspace{12pt}

(A) infinite coordination ($z\rightarrow\infty$) (Dobrosavljevi\'{c} \&
Kotliar 1993, 1994), or\vspace{6pt}

(B) non-interacting electrons for arbitrary coordination
(Anderson 1958).\vspace{12pt}

Note that in the case (A) the sum in the definition of $\Delta_{i}
(\omega_n )$ runs over infinitely many neighboring sites, so that
the hybridization function is replaced by its {\em average value}.
As a result, the all the spatial fluctuations in the ``cavity''
representing the environment of a given site are suppressed,
prohibiting Anderson localization effects. For noninteracting 
electrons, the resulting theory reduces to the well known
``coherent potential approximation'' (Elliott et al. 1974).

Since one of the main goals of the statistical DMF theory is to 
incorporate Anderson localization effects,   
we concentrate on finite coordination lattices.
In this case, $\Delta_{i}$ can be considered to be 
a {\em functional} of the lattice Green's functions 
$G_{jk}$, evaluated for fixed disorder, i. e.

\begin{equation}
\Delta_i =\Delta_i \; [G_{jk}].
\end{equation}
At strong disorder, $\Delta_{i}$ will exhibit pronounced
fluctuations from site to site, reflecting a distribution 
of local environments ``seen''  by the electrons. 

Finally, in order to obtain a closed set of DMF self-consistency
conditions, we need to specify a procedure that relates the lattice Green's
functions $G_{jk}$ to the solution of the local dynamical
problem, as  defined by the local effective action $S_{eff} (i)$. We first
note that the local action $S_{eff} (i)$ is identical to that
describing an Anderson impurity (AI) model (Anderson 1961, Hewson
1993) embedded in a sea of conduction electrons described by a
hybridization function $\Delta_{i}(\omega_n )$. The solution of this
AI model then uniquely defines the corresponding self energy 
$\Sigma_i$, which is given by
\begin{equation}
\Sigma_i(\omega_n )= 
i\omega_n +\mu -\varepsilon_i -\Delta_i (\omega_n )
-(G_{ii}^{loc} (\omega_n ))^{-1},
\end{equation}
where the local Green's function 
\begin{equation}
G_{ii}^{loc} (\omega_n ) = <c^{\dagger}_{i}(\omega_n )c_{i} (\omega_n )>_{loc},
\end{equation}
is evaluated with respect to the local effective action $S_{eff} (i)$. 

Next, we follow an ``exact eigenstate'' 
strategy, and define the ``bare'' lattice Green's functions
$G_{ij}^o$ as the exact lattice Green's functions evaluated 
for the same realization of disorder $\{\varepsilon_i \}$, in 
absence of interactions. The ``full'', i. e. interaction-renormalized
lattice Green's functions are within statistical DMF theory then defined by
\begin{equation}
G_{ij}=G_{ij}^o [\varepsilon_i\rightarrow\varepsilon_i + 
\Sigma_i(\omega_n )],
\end{equation}
closing the set of DMF self-consistency conditions. 

We emphasize that a similar relationship relates the {\em exact} 
interaction-renormalized lattice Green's functions to their 
noninteracting counterparts. However, in the exact formulation, 
the self energies describing the interaction renormalizations
are in general {\em nonlocal in space}, as well as in frequency. 
Having this in mind, we can describe the statistical DMF theory
as a requirement for these interaction dependent self energies 
to assume a strictly {\em local} character. 

For an arbitrary lattice, an iterative procedure for solving 
the above set of statistical DMF equations could be obtained by
\vspace{12pt}

(i) making an initial guess for the form of 
$\Delta_i (\omega_n )$, 

(ii) solving the corresponding AI models on every lattice site,

(iii) use the resulting $\Sigma_i$-s to calculate the full lattice
Green's functions,

(iv) calculate the new values of $\Delta_i (\omega_n )$
and go back to step (ii).\vspace{12pt}

This procedure can be carried out for an arbitrary lattice in any
dimension, but the procedure can be fairly time consuming due to the
need to compute all the off-diagonal components of the lattice Green's
functions $G_{ij}$.  However, similar computations have already been
carried out to study the interplay of correlations and disorder in a
Hartree-Fock (HF) approach (Yang \& MacDonald 1993).  Interestingly, such a
lattice HF treatment for a fixed disorder realization can be obtained
as a further simplification of the statistical DMF theory, if the
local Anderson impurity models are themselves solved at the same
level. Of course, an exact solution of the statistical DMF equations
goes well beyond such a HF treatment, since it can describe {\em
inelastic} processes, which are expected to be of particular
importance at finite temperature and in metallic non Fermi liquid
phases.

In practical terms, it is important to identify specific models
where the solution of the statistical DMF equations can be
simplified.  The situation is particularly simple in the case of a
Bethe lattice (Cayley tree). Because the absence of loops on such
lattices, only {\em local} Green's functions $G_{jj}^{(i)} (\omega_n )$
appear in the expression for $\Delta_i (\omega_n )$.  Furthermore, in
this case the objects $G_{jj}^{(i)} (\omega_n )$ can be computed from a
local action of the form {\em identical} as in Eq. (5.2), {\em except}
that in the expression for $\Delta_{j} (\omega_n )$, the sum now
runs over $z-1$ neighbors, excluding the site $i$.
We thus conclude that the objects $G_{jj}^{(i)} (\omega_n )$ are 
related by a {\em stochastic} recursion relation, that involves 
solving Anderson impurity models with random on-site energies 
$\varepsilon_i$. In the non-interacting limit, the recursion 
relations can be written in close form, and reduce 
to the exact solution of the disordered electrons 
on a Bethe lattice (Abou-Chacra et al. 1971). 

\section{Order parameters  for the metal to insulator transition:
 the disordered case}
\label{new}

In attempting to describe any phase transition, 
a crucial step is to identify  appropriate
{\em order parameters} that can characterize the qualitative 
differences between the different phases 
of the system. 

Since the basic focus of the dynamical mean
field approach
is the one particle Green's function,
it is useful to write down an explicit expression
for the Greens function of a lattice system with a 
one body Hamiltonian $H_{ij}$ and a general two body interaction
term in terms of a self energy $\Sigma_{ij} (i \omega)$.
\begin{equation}
[G^{-1} (\omega_n)]_{ij} =[\delta_{ij} i \omega_n - H_{ij} 
- \Sigma_{ij} (\omega_n )].
\end{equation}
Here, we have used a matrix notation, for a fixed realization of disorder, 
so that the self energy $\Sigma_{ij} (\omega)$ describes the interaction
induced renormalizations of the Green's function. 
A quasiparticle picture emerges under a fairly general assumptions of
regularity of the interaction self energy at low frequencies.
In this case
\begin{equation}
\Sigma_{ij} ( \omega) \approx 
\Sigma_{ij} (0) +([Z^{-1}]_{ij}-1) \omega_n + O(\omega_n^2).
\end{equation}
and the Greens function at low frequencies assumes a very transparent form
\begin{equation}
G (\omega_n )= \sqrt{Z} G^{QP} (\omega_n )\sqrt{Z}
\end{equation}
with the quasiparticle Green's function defined
as
\begin{equation}
[G^{QP}]^{-1}= \omega_n I - \sqrt{Z} 
(H+\Sigma(0)) \sqrt{Z} .
\end{equation}
Introducing the low energy eigenvalues of the quasiparticle Hamiltonian
$E_{n}$ and their corresponding eigenvector $|n>$, we can write
\begin{equation}
G(x,x') \approx \sum_{n} {{< x | \sqrt Z | n> < n  | \sqrt Z | x'> }\over
{ ( \omega - E_{n})}}.
\end{equation}

So far we have just repeated the derivation of the disordered
Fermi liquid framework of dirty metals ( Castellani et al. 1987).
In that reference those concepts were used to interpret the
field theoretical coupling constants of
Finkelshtein's non linear sigma model
in terms of Fermi liquid parameters.
Implicit in that  framework was the assumption
that the spatial fluctuations, or the sample to sample fluctuations
were not too large. This  assumption is valid for weak disorder.
The main advance of the dynamical mean field approach is
the  ability to deal with strong spatial fluctuations.

The results of the next section indicate very different behavior for
typical and average quantities, reflecting large spatial fluctuations.
These differences may be the source of the finite scale divergences that were 
encountered in the field theoretical approach (Finkhelstein 1983, 1984)
which, from the very begining,  carries out an 
average over the disorder.

We now use the previous developments to introduce the order parameters
which are relevant to the DMF theory.  We begin with defining 
the local density of states at zero frequency
\begin{equation}
\rho_{i} = -{ 1 \over \pi} Im G_{ii} (0+),
\end{equation}
and the corresponding local quasiparticle density of states
\begin{equation}
\rho^{QP}_{i} = -{1 \over \pi} Im G^{QP}_{ii} (0+).
\end{equation}
Their ratio defines the second order parameter 
\begin{equation}
Z_{i} = {\rho_{i} \over \rho^{QP}_i}.
\end{equation}
which could be dubbed the local quasiparticle weight. 
In the DMF framework, all the previous equations simplify because the
self energy is local ($\Sigma_{ij} ( \omega_n)=\delta_{ij}\Sigma_{i} (
\omega_n)$) and $Z_{i}$ can directly be calculated from the expression
\begin{equation}
{Z_i}= \left[1-{\partial \over {\partial  \omega} 
}Re[\Sigma_i (\omega_n )]|_{\omega_n =0+}\right]^{-1}.
\end{equation}

When these parameters are mostly uniform, i.e. site independent
their averages  are precisely two of the coupling constants in the field
theoretical approach of Finkelshtein. In the statistical mean field
approach we focus instead on whole distributions (or on typical values) of
these parameters.

We now turn to the physical interpretation of the order parameters, or
more precisely their statistical distribution.  The low energy physics
is described by two parameters, $\rho_i$ and $Z_i$, which are
associated with the height and the width of the resonance in the
spectra of the local impurity problem. Physically, $\rho_i$ can be
interpreted as a the density of states for adding or removing and
electron from a specific site; $Z_i$ is related to the energy scale
(or timescale) over which the quasiparticle picture applies. On the
metallic side of the transition, the electron behaves as a local
magnetic moment up to a time scale proportional $Z_i ^{-1}$, but as a
delocalized quasiparticle over longer timescales.  The vanishing of
$Z_i$ signals the conversion of quasiparticle degrees of freedom into
local moments, which takes place at the Mott transition.
 
To gain further insight into the physical content of these order
parameters, we concentrate on transport properties. The fundamental
difference between a metal and an insulator is defined by  the 
ability of the electron to leave a given lattice site -- to delocalize. 
In early work (Anderson 1958) Anderson pointed out that, at least 
for noninteracting electrons, this property could be examined by evaluating
the {\em lifetime} of an electron on a given site.
Very generally,  the inverse lifetime simply measures the width of the 
the local resonance level. We thus expect
\begin{equation}
\frac{1}{\tau_i }\sim \Im (G_{ii} (0+))^{-1}.
\end{equation}
Within DMF, the local Green's function takes the form
\begin{equation}
G_{ii} (\omega_n) = [i\omega_n +\mu -\varepsilon_i -\Delta_i (\omega_n )
-\Sigma_i(\omega_n )]^{-1},
\end{equation}
and since for  Anderson impurity models (see for example Hewson 1993)
$
\Im \Sigma_i (0+)=0,
$
we conclude that the desired local lifetime can be directly related to
the local hybridization function as

\begin{equation}
\frac{1}{\tau_i }\sim \Im \Delta_i (0+).
\end{equation}

We thus expect $\Im \Delta_i (0)$  to vanish whenever the system is
insulating, but to remain finite in a metallic regime.  We
emphasizing that the {\em same} qualitative behavior should be expected
for the local density of states (LDOS) $\rho_i$ as can be readily seen
from Eq. (5.3). We can thus use the LDOS as an order parameter that
discriminates a metal from an insulator.  In a random system, $\rho_i$
will fluctuate from site to site, and we need a whole distribution
function to fully characterize the approach to the transition.

In particular, in the  Mott insulator, there is a
``hard'' gap of order $U$ on every lattice site, so even the {\em
average} DOS discriminates the Mott insulator from the metal. The
situation is more complex as the Anderson insulator is
approached. Here, the local spectrum is composed of a  few delta-functions
(discrete, bound states), separated by gaps, but the average DOS
remains {\em finite}.  In contrast to the Mott insulator, in the
Anderson insulator the sizes and positions of the local gaps {\em
fluctuate}, but in both cases a {\em typical} site has a gap at the
Fermi energy.  A natural order parameter is therefore the {\em
typical} DOS, that is represented by the {\em geometric average}
\begin{equation}
\rho_{typ}=<\rho >_{geom} =\exp\{ <\ln \rho_i >\}. 
\end{equation}
This quantity is found
to {\em vanish} at the Anderson transition, in contrast to the 
{\em average} DOS, which is not critical. 

On the metallic side of the transition,  the distribution
function of a second quantity, the local quasiparticle (QP)
weight, is necessary to characterized  the low energy behavior near
the transition. Important information is obtained from
the typical value of   the  random variable $Z_i$, 
 defined as
\begin{equation}
Z_{typ}=\exp\{<\ln Z_i >\},
\end{equation}
which emerges as a natural order parameter from previous
studies of the Mott transition.

Finally, we  define the {\em averaged} QP DOS by
\begin{equation}
\rho^{QP}_{av}= <\rho^{QP}_i >.
\end{equation}
This 
object is very important for thermodynamics, since it is
directly related to quantities such as the specific heat
coefficient $\gamma =C/T$, or the local spin susceptibility $\chi_{loc}$. 
Note that in absence of interactions, $\rho^{QP}_{av}$ 
reduces to the usual (algebraic) average DOS, which is not
critical at a $U=0$ Anderson transition, but it is strongly enhanced
in the vicinity of the Mott transition.

It is instructive to discuss the behavior of these order parameters in
the previously studied limiting cases.  In the limit of large lattice
coordination, the spatial fluctuations of the bath function $\Delta_i
(\omega_n )$ are unimportant, and there is no qualitative difference
between typical and average quantities.  In the Mott insulating phase
of a periodic solid, there is a gap in the density of states,
while there is a finite
density of states on the metallic side of the transition.  As the MIT
is approached from the metallic side, $\rho_{typ}$ remains finite, but
$ Z_{typ}$ is found ( Dobrosavljevi\'{c} \& Kotliar 1993,1994) to
linearly go to zero.

Another well studied limit is that of noninteracting electrons on the
Bethe lattice, which is known 
(Abou-Chacra et al. 1971, Efetov 1987, Mirlin \& Fyodorov 1991)
to display an Anderson transition.  The average DOS is {\em finite}
both in the insulating and in the metallic phase, and is non critical
at the transition.  Similarly, by definition $ Z_{typ}=1$ in this
noninteracting limit, so it also remains non critical.  On the other
hand, the typical density of states $\rho_{typ}$ is finite in the
metal and zero in the Anderson insulator.  This quantity is critical,
and is found to vanishes exponentially  with the distance
to the transition.

The definitions of the the
order parameters that we have proposed are {\em not restricted} to the
statistical DMF framework, which is simply used as a specific 
calculational scheme. The same definitions can in principle be used
in other approaches that can calculate local {\em unaveraged}
values of the local DOS $\rho_i (\omega_n )$ or the local part of the 
frequency dependent self energy $\Sigma_{ii} (\omega_n )$ due to the
interactions.

\section{Results}

The dynamical mean field theory maps the insulating phase
of the model onto a collection of Anderson impurity models
each one of them is embedded in an insulating bath.
An  Anderson impurity in an insulator, away from particle
hole symmetry,  can either  have a 
doublet  ground state when the coupling to the enviroment is weak
or a singlet ground state when the coupling to the bath exceeds
a critical value.
In this  strong coupling limit
(Dobrosavljevi\'{c} \&
Kotliar 1992$b$), a {\em bound state}  is pulled from the continuum
formed by the  bands of the insulator.
{}From these general consideration,  we obtain a two fluid
picture of the insulating phase:  there are sites which have
a local moment down to zero temperature while other sites
quenched their spin by exchanging it with a strongly coupled neighbor.

The number of sites in a doublet state, in the insulating
phase of the system form   ``Mott
droplets" . 
Based on our experience on the Mott transition in clean
systems, we expect 
a larger
density of states  when the system
undergoes a metal to insulator transition into an insulating phase
with a large number of doublets. 
This result in   at least two different regimes
depending on whether the insulating phase  has
a 
high or low concentration of sites in  doublet states.

We have considered a $z=3$
(Dobrosavljevi\'{c} \& Kotliar 1997). 
Bethe lattice, in the limit of infinite on-site repulsion $U$ at $T=0$
and fixed average density $n$ in the presence of a uniform
distribution of random site energies $\varepsilon_i$ of width $W$. To
calculate the probability distributions of $\rho_j$ and $Z_j$ we used a
simulation approach, where the probability distribution for the
stochastic quantity $G_{j}^{(i)} (\omega_n )$ is sampled from an
ensemble of N sites, as originally suggested by Abou-Chacra et
al. (1971).  In order to solve Anderson impurity models for given bath
functions $\Delta_{j} (\omega_n )$ we use the slave boson (SB)
mean-field theory (Barnes 1977, Read \& Newns 1983, Coleman 1987),
which is known to be qualitatively and even semiquantitatively correct at
low temperature and at low energies. In agreement with  heuristic
arguments, we expect the results to be a strong function of the
density $n$.  In order to illustrate this behavior, we have
carried out explicit calculations  for both low electron density 
$n=0.3$, and high electron density  $n=0.7$ (i.e. close to half filling).
We emphasize that in the
clean limit ($W=0$) the behavior is qualitatively identical for the
two values of the density, and the system remains metallic, with only the value of
the effective quasiparticle mass $m^*\sim Z^{-1}$ being a function of
$n$.  Qualitatively different behavior is found as the disorder is
introduced.

We first describe the evolution of the probability
distribution of the local quasiparticle weights $Z_i$, as the disorder
is increased.  The sites with $Z_i \ll 1$ represent (Milovanovi\'{c}
et al. 1989, Dobrosavljevi\'{c} \& Kotliar 1993,1994) disorder-induced
local magnetic moments, and as such will dominate the thermodynamic
response (see the definition of $\rho_{QP}$).  For weak disorder we
expect relatively few local moments and the quasiparticle weight
distribution is peaked at a finite value.  As the disorder is
increased, the distribution of $Z_i$-s broadens.  At a critical value
of the disorder $W=W_{nfl}$, the form of this distribution assumes a
singular form (Bhatt \& Fisher 1992, Dobrosavljevi\'{c} et al. 
1992$a$, Miranda et al. 1996, 1997) , leading to anomalous thermodynamic
response characterized by a diverging magnetic susceptibility $\chi$
and specific heat coefficient $\gamma$.  This behavior was found
both for high and low density, in remarkable agreement with
experiments carried out on for uncompensated and compensated doped
semiconductors.  Interestingly, a similar transition to a non Fermi
liquid metallic phase, well before the MIT, has been found from the
field-theoretical approaches in $2+\varepsilon$ dimensions
(Finkelshtein 1983,1984; for a review of subsequent work on this
issue, see the review by Belitz \& Kirkpatrick (1994)).

As the level of disorder is increased further, a metal insulator
transition is reached at the second critical value of disorder
$W=W_c$. At this point, the typical quasiparticle weight $Z_{typ}$
vanishes, a behavior reminiscent of the clean Mott transition. While
this behavior is found at both high and low density, the examination
of the second order parameter $\rho_i$  reveals striking density
dependence in the vicinity of the transition. The difference are most
clearly displayed by plotting the evolution of the {\em distribution
function} $P(\rho_i )$ as a function of doping. 
\begin{figure}
\centerline{
\epsfxsize=3.2in \epsfbox{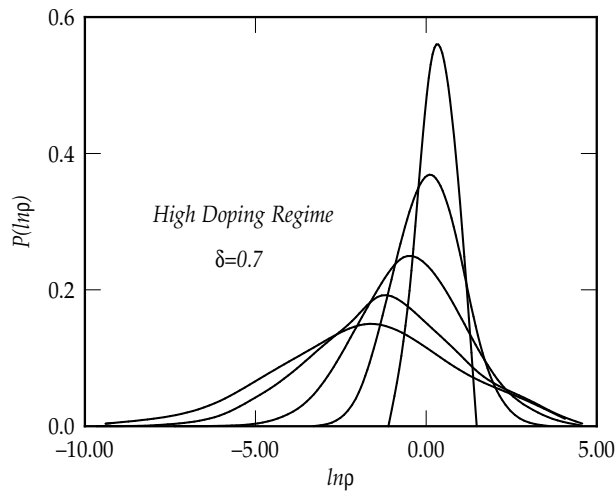}\vspace{6pt}}
\caption{DOS statistics in the low filling (high doping) regime.  The full
distributions $P(\ln \rho )$ are presented for increasing amounts of
disorder. We find that the {\em maximum}, i. e. $<\ln\rho >$ shifts,
as the transition is approached. Note also the extremely large {\em
width} of the distribution, so that $\rho$ now spans many orders of
magnitude.}
\end{figure}
At low filling
$n=0.3$, we find that the with of this distribution becomes extremely
large, spanning many decades, reflecting huge spatial fluctuations of
the electronic wave-functions, as shown in Fig. 1.  At the same time,
the most probable value $\rho_{typ}$ is found to dramatically
decrease, vanishing linearly with the distance to the transition.  In
contrast to this result, a very different behavior is found at high
density $n=0.7$. 
\begin{figure}\centerline{
\epsfxsize=3.2in \epsfbox{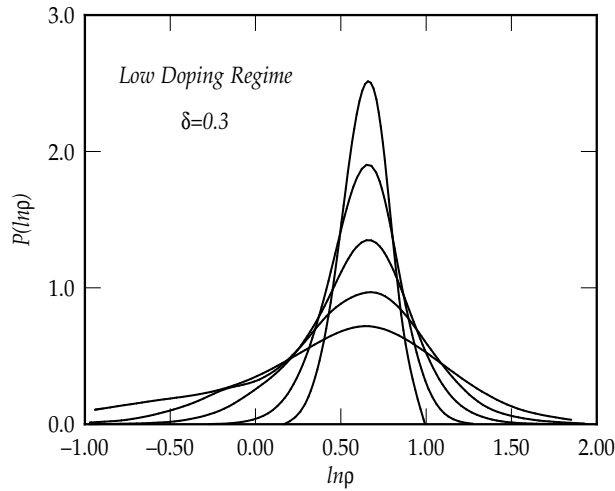}\vspace{6pt}}
\caption{DOS statistics in the high filling (low doping) regime. 
W show the probability distribution for $\ln \rho$, as
a function of disorder. 
We clearly see that while the 
distribution  broadens, $<\ln q>\approx cons$. 
\label{fig2}}
\end{figure}
As we can see from Fig. 2, in this case the
distribution {\em width} again broadens, albeit in a somewhat slower
fashion, as the transition is approached.  However, in this case the
most probable value $\rho_{typ}$ is only weakly modified with
increasing disorder, and is found to approach a {\em finite} value at
the transition. Recalling that the statistics of $\rho_i$ are
intimately related to {\em transport}, our results strongly suggest
that kinetic coefficients such as the conductivity will strongly
depend electron filling, as the MIT is approached. These expectations
should be confirmed by explicitly calculating the critical behavior
conductivity at different values of the density.

Finally, we mention an important feature of the insulating state in
presence of both the strong electronic correlation and disorder.  By
explicit calculations of both the average and the typical density of
states we have demonstrated that 
the introduction of disorder, 
fills
the Mott gap 
with localized electronic states. The average DOS is therefore 
finite  in the insulator. Nevertheless, the MIT still retains a
definite Mott character, with  a finite fraction of electrons turning
into localized magnetic moments, and the typical quasiparticle weight
$Z_{typ}$ vanishing at the transition. 
Our
results explicitly show that the metal insulator transition in
presence of both interactions and disorder is a qualitatively new type
of transition, having well defined signatures of {\em both } the Anderson and
the Mott route to localization.

\section{Conclusions}
\label{conclusions}

The metal to insulator transition  continues to be one of the central
problem in condensed matter physics.  Building on the fundamental
concepts introduced by Mott and Anderson there have been several
attempts at the construction of a microscopic theory.  We have
summarized in this contribution some aspects of the dynamical mean
field approach to this problem.

Deeply rooted in local physics, it combines the  chemical
and physical aspects  of the problem. We regard the
construction of a mean field theory as an essential first step
before a more complete treatment including Gaussian and
non linear fluctuations is carried out.
In the clan limit the DMF description of the Mott transition
has already
given insights into puzzling aspects of transition metal oxide physics.

Since the approach emphasizes the local environment, it has a 
conceptually very simple extension to the disorder case: the statistical
dynamical mean field theory. The theory explicitly incorporates both
the Mott and the Anderson route to localization, and thus provides a
consistent description of the transition, interpolating between the
respective limits of no disorder and no interaction.  This  key
feature seems to be missing in a recent  field theoretical formulation of the
Anderson-Mott transition (Kirkpatrick \& Belitz 1994, 1995).

The statistical mean field theory reproduces many remarkable features
of doped semiconductors. A transition to a Griffiths phase 
(Bhatt and Fisher 1992, Dobrosavlevic et. al. 1992, Lakner et. al. 1994)
where local
moments coexist with conduction electrons down to arbitrary low
temperature precedes the true metal to insulator transition. The
local properties such as the density of states depend in a distinct
fashion on the level of doping (compensation).  The metal to insulator
transition in the low doping region (uncompensated) has a larger
typical density of states and is closer to a Mott transition than the
corresponding transition at larger doping levels (compensated
situation).  The latter contains ingredients from both the Anderson
and the Mott transition.

There are several issues that deserve further investigation.  A
detailed study of the transport properties has to be carried out.  One
should also elucidate the interplay of the short wavelength
fluctuations with long wavelength modes.  The latter are presumably
described by the non linear sigma model approaches that have been
investigated near two and six dimensions.  
A second element still missing form the mean field theory
is the inclusion of short range magnetic
correlations.
In our view these are the most pressing problems that need to be address
on the road to a  comprehensive theory of the Mott-Anderson transition.

\pagebreak
\vspace{24pt} 

\noindent {\bf  Acknowledgements}: 

\vspace{24pt} 
We are grateful to E. Abrahams, R. Bhatt, N. Bonesteel, L. Gorkov, E. Miranda, 
D. Popovi\'{c}, M. P. Sarachik, J. R. Schrieffer, and G. Thomas for useful
discussions. VD was supported by the
National High Magnetic Field Laboratory at Florida State University,
and the Alfred P. Sloan Foundation.
GK was supported by NSF DMR 95-29138.

\end{document}